\documentclass[12pt]{article}
\usepackage{graphicx}
\title{Hadron production in lepton-nuclei
 interactions at high energies:\\ Monte Carlo generator HARDPING 2.0}

\author{Ya.\,A.\,Berdnikov$^{\&\%}$, A.\,E.\,Ivanov$^{\&\%}$\/\thanks{e-mail:
ivanovae@pnpi.spb.ru}, \,V.\,T.\,Kim$^{\&\%}$, V.\,A.\,Murzin$^{\%}$\\~\\
$^\&$St. Petersburg State Polytechnical University, 195251 St. Petersburg, Russia\\
$^\%$St. Petersburg Nuclear Physics Institute NRC KI, 188300 Gatchina, Russia}

\begin{document}

\maketitle

\begin{abstract}
Hadron production in lepton-nucleus interactions at high-energies is considered
in framework of developing Monte Carlo event generator HARDPING (HARD Probe
INteraction Generator).  Such effects as formation length, energy loss and
multiple rescattering for produced hadrons and their constituents are implemented into the HARDPING 2.0. 
Available data  from HERMES collaboration on hadron production in lepton-nucleus collisions 
are described by the present version of the HARDPING generator in a reasonable
agreement.
\end{abstract}

Hadronisation of quarks and gluons is one of the most intriguing parts of
nonperturbative QCD. Use of nuclear targets may allow to reveal important
features of space-time picture of hadronisation, like hadron formation length
and energy loss, see, e.g, for a review \cite{Kopeliovich:2003py,Baier:2000mf}
and references therein.
The understanding of quark propagation in nuclear medium is crucial for the
interpretation of ultrarelativistic heavy ion collisions, as well as high energy
proton-nucleus and lepton-nucleus interactions. To simplify interpretation of
observable effects one can consider at the beginning hadron production in 
lepton scattering off nuclei. 
In case of deep inelastic  scattering of lepton on nucleus there can be two
stages of hadronisation. The first stage is predominantly perturbative. At this stage after hard scattering a struck
a struck quark propagates through the nuclear medium being in point-like parton state
experiencing a little attenuation only. This effect is known as 
 Landau-Pomeranchuck-Migdal effect in QCD
\cite{Baier:1996sk,Zakharov:1997uu,Levin:1995mx,Wiedemann:2000za,Wang:2001ifa,
Gyulassy:2000er, Arnold:2002ja,Zapp:2008af}. At the end of the first stage, a pre-hadron
state (a color dipole or constituent quark) is formed
\cite{Accardi:2002tv,Accardi:2005jd,Domdey:2008aq}. 
In the second stage pre-hadron state with smaller than hadron cross section interact with nuclear medium. There is finally
formed hadron at the second stage. 
At the large enough energies of  produced hadrons the nonperturbative stage of hadron formation 
is evolving beyond the nucleus \cite{Benhar:1997zz}.

The aim of the work is to study these effects for case of lepton-nuclei
collisions using a developing Monte Carlo (MC) event generator. The generator
HARDPING (HARD Probe INteraction Generator) is based on MC generators
PYTHIA \cite{Sjostrand:2006za} and HIJING \cite{Gyulassy:1994ew}. 
The first version of HARDPING describes experimental data on Drell-Yan reaction
off nuclei reasonably well \cite{Berdnikov:2005vh,Berdnikov:2006kt}. It takes
into account the effects related with interaction of projectile hadron and its
constituents in nuclear matter  before hard scattering for lepton-pair
production off nuclei. 
The second version of HARDPING, presented here, describes, in addition, hadron
production in lepton-nuclei interactions. It incorporates the following effects:
formation length, energy loss and multiple soft rescatterings.

The experimental results on semi-inclusive leptoproduction of hadrons off nuclei
\cite{Airapetian:2003mi,Airapetian:2011jp} are presented in terms of hadron
multiplicity ratios \(R^{h}_{M}\) with nuclear ($A$) and deuteron ($D$) targets, as
functions of virtual photon energy (\(\nu\)),  its fraction taken by hadron (\(z_{h}\))
 and
hadron transverse momentum squared \(p_{\perp}^{2}\):
\begin{equation}
R^{h}_{M}\left(x\right)=\frac{1}{N_{A}^{DIS}}\frac{dN^{h}_{A}}{dx}\Big/\frac{1}{
N_{D}^{DIS}}\frac{dN^{h}_{A}}{dx}
\end{equation}
where \(N^{DIS}_{A}\) and \(N^{DIS}_{D}\) are yelds of inclusive deep-inelastic
scattering leptons on nuclei $A$ and $D$, \(\frac{dN^{h}_{A}}{dx}\) and
\(\frac{dN^{h}_{D}}{dx}\) are yields of semi-inclusive hadrons as a function of
$x$, here $x$ is either $z_{h}$ or \(p{\perp}^{2}\).
In absence of nuclear effects, the ratio \(R^{h}_{M}\) should be equal to 1. The
experimental results show that this is the case at high transferred energy
\(\nu\) \cite{Airapetian:2003mi}.

It is well established from theoretical and experimental studies of
hadron-nucleus collisions at high energy that hadrons are not produced at the
point of collision but only after some ``formation'' length
\cite{Kopeliovich:2003py}.
In the Lund string fragmentation model, the production of hadrons is described as
two stage process. At the first perturbative stage a pre-hadron at the end of
the string is formed. On the next nonperturbative stage a hadron is formed.
Before a pre-hadron is formed, the struck quark propagates through the nuclear
matter with a very small cross section (in this work we neglect it). It takes
some time at the perturbative stage to form a pre-hadron 
(formation time, $t_p$ or formation length, $l_p$). 
When the pre-hadron is formed, it interact with nuclear matter via pre-hadron cross
section, which is different from hadronic cross section. And also it takes an
extra time to form the final hadron from pre-hadron state. So, the formation length consist
of two parts (\(l_{p}\) and \(l_{n}\)), corresponding the two stages of
hadronisation.

There are two approaches to calculate formation length with two stages. The first
is based on an oversimplified description of nonperturbative stage 
\cite{Kopeliovich:2003py}, while the second one \cite{Accardi:2002tv} is based
on Lund string model, but neglecting the energy loss effects during the perturbative stage.
The present work is based on the both above approaches with including
the effect of energy loss at the perturbative stage and using Lund string model 
at the nonperturbative stage. 

\begin{figure*}[htb]
\centering
\vspace*{0.0cm}
\includegraphics[width=0.9\textwidth, height=0.30\textheight]{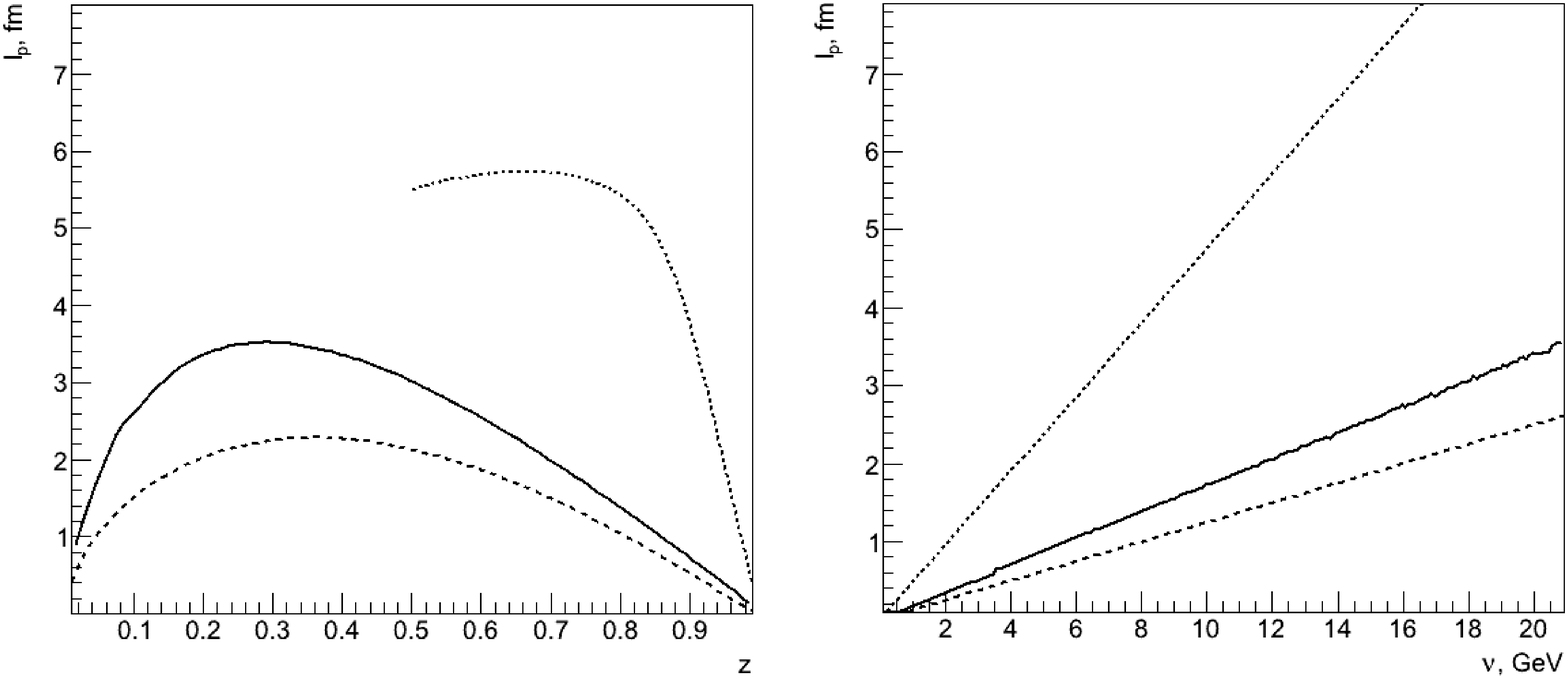}
\caption{Formation length as a function of \(z_{h}\) and as a function of \(\nu\). The dotted
lines correspond to the first approach (B.Z. Kopeliovich {et al.}
\cite{Kopeliovich:2003py}) for $z_{h}$ \(>\) 0.5, the dashed lines corresponds to the second
approach (A. Accardi et al. \cite{Accardi:2002tv}) and the solid lines corresponds to
HARDPING 2.0 calculations}
\label{fig:lp}
\end{figure*}

In the first approach the distribution on pre-hadron formation time can be written
in the following form \cite{Kopeliovich:2003py}:
\begin{eqnarray}
&
W\left(t_{p},z_{h},Q^{2},\nu\right)=N\int\limits^{1}_{0}\frac{d\alpha}{\alpha}
\delta\left[z_{h}-\left(1-\frac{\alpha}{2}\right)\frac{E_{q}\left(t_{p}\right)}{\nu}
\right]\times {} \nonumber\\
&
\int\limits^{Q^{2}}_{\Lambda_{QCD}^{2}}\frac{dk_{\perp}^{2}}{k_{\perp}^{2}}\delta\left[k_{\perp}^{
2}-\frac{2\nu}{t_{p}}\alpha\left(1-\alpha\right)\right]\int{}dl_{\perp}^{2}\delta\left[
l_{\perp}^{2}-\frac{9}{16}k_{\perp}^{2}\right]\times {}\nonumber\\
& 
\int\limits_{0}^{1}d\beta\delta\left[\beta-\frac{\alpha}{2-\alpha}\right]
\left|\Psi_{h}\left(\beta,l_{\perp}\right)\right|^{2}S\left(t_{p},z_{h},Q^{2},\nu\right)
\end{eqnarray}
here $t_{p}$ is pre-hadron formation time, $z_{h}$ is fraction of virtual photon
energy carried out by the hadron, $Q^{2}$ is virtual photon virtuality, $\nu$ is
virtual photon energy, \(\Lambda_{QCD}\) is QCD constant, \(\Psi_{h}\left(\beta,l_{\perp}\right)\) 
meson wave function, $E_{q}\left(t_{p}\right)=\nu-\Delta E\left(t_{p}\right)$ is quark
energy and \(\Delta{}E\left(t_{p}\right)\) is quark energy losses 
due to perturbation gluon radiation,
 \(\delta\) is delta-function, \(S\left(t_{p},z_{h},Q^{2},\nu\right)\) is Sudakov suppression factor.
$\Delta E\left(t_{p}\right)$ can be taken in the following form:

\begin{eqnarray}
&
\Delta{}E\left(t\right)=\nu{}\int\limits^{Q^{2}}_{\Lambda_{QCD}^{2}}dk_{\perp}^{2}\frac{4\alpha_{s}\left(k_{\perp}^{2}\right)}{3\pi}\int\limits^{1}_{0}d\alpha{}\frac{1}{k_{\perp}^{2}}\times {} \nonumber\\
&
\theta\left(t-\frac{2\nu\alpha\left(1-\alpha\right)}{k_{\perp}^{2}}\right)\theta\left(1-z_{h}-\alpha\right)
\end{eqnarray}
where \(\theta\) is step-function, \(\alpha_{S}\) is strong coupling.
This approach works only for leading hadrons with $z_{h}$ \(>\) 0.5, (see Fig.\ref{fig:lp}, the dashed line), which is not
suitable for full MC simulation.

The second approach is based on Lund string model for nonperturbative hadronisation 
neglecting its perturbative stage  \cite{Accardi:2002tv}. 
In this approach probability to have pre-hadron formation length $l_{p}$ can be written
in the following form:
\begin{eqnarray}
& P\left(l_{p};z_{h},L\right)=
\frac{z_{h}L}{l_{p}-z_{h}L}\left[\frac{l_{p}}{\left(l_{p}+z_{h}L\right)\left(1-z_{h}\right)}\right]^{C}
\times {}\nonumber\\
&
\left(\delta\left[l_{p}-\left(1-z_{h}\right)L\right]+\frac{1+C}{l_{p}-z_{h}L}\theta\left[
\left(1-z_{h}\right)L-l_{p}\right]\right) {}\\
& \times\theta\left[l_{p}\right] {}. \nonumber
\end{eqnarray}
where parameter $C$=0.3 \cite{Accardi:2002tv}, and parameters
\(k\) and $L$ are string tension and ratio of the virtual photon energy to
string tension \(L={\nu}/{k}\).

In the presented here approach the effect of energy loss was incorporated into the HARDPING 2.0 
using PYTHIA MC implementation of parton shower for the perturbative stage and Lund string model for the nonperturbative one.
On the Fig.\ref{fig:lp} the dependences of formation length on $z_{h}$ and
\(\nu\) are plotted, the dotted lines - the first approach \cite{Kopeliovich:2003py},
the dashed lines - the second approach \cite{Accardi:2002tv}, the solid lines - HARDPING 2.0 simulation.

During the perturbative stage, corresponding to the formation length \(l_{p}\), a constituent quark  (or pre-hadron)  
state is formed. It can interact with intranuclear nucleons via
inelastic pre-hadronic cross-section
 (or inelastic  quark-nucleon cross-section). At the end of the nonperturbative stage the observed
hadron is formed.

Produced pre-hadrons and hadrons can undergo soft collisions with intranuclear
nucleons (with small momentum transfers: \(|t| < \) 1 \(GeV^{2})\). 
So, one has to take into account
their soft multiple rescattering.

The transverse momentum distribution of constituent quarks after one soft interaction can be
parameterised in the following form
\cite{Efremov:1985cu,Berdnikov:2005vh,Berdnikov:2006kt}:
\begin{equation}
f_{p}\left(\vec{p_{\perp}}\right)=\frac{B^{2}}{2\pi}e^{-Bp_{\perp}}
\end{equation}
where \(B=\frac{2}{\left<k_{p}\right>}\), where $ \left<k_{p}\right> $ is mean
value of quark transverse momentum.
\(f_{p}\left(\vec{p_{\perp}}\right)\) is a differential distribution of quark in 
 quark-nucleon interaction normalised on unity.

Probability to have no interactions between the points with coordinates 
\(\left(z,\vec{b}\right)\) and \(\left(z+\lambda,\vec{b}\right)\) can be written in the next form:
\begin{equation}
P\left(\lambda{};z,\vec{b}\right)= e^{-\sigma T\left(\vec{b},z,\lambda\right)}
\end{equation}
where \(T\left(\vec{b},z,\lambda\right)\) is:
\begin{equation}
T\left(\vec{b},z,\lambda\right)=\left(A-1\right)\int^{z+\lambda}_{z}{
\rho\left(\vec{b},z^{'}\right)dz^{'}},
\end{equation}
\(\rho\left(\vec{b},z\right)\) is nuclear density and \(\sigma\) is quark-nucleon 
(pre-hadron-nucleon) or hadron-nucleon inelastic cross section.

\begin{figure*}[htb!]
\centering
\vspace*{0.0cm}
\includegraphics[width=0.9\textwidth, height=0.3\textheight]{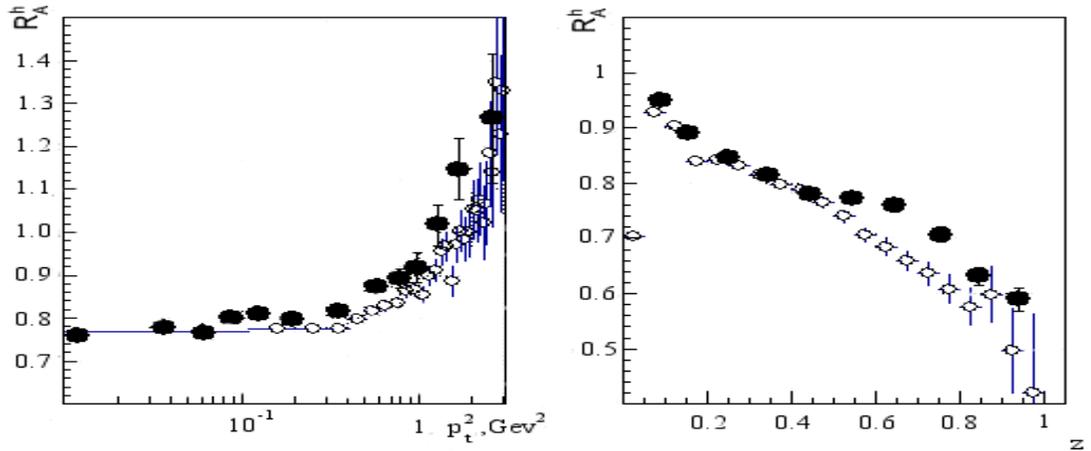}
\caption{Multiplicity ratio (\(R_{M}^{h}\)) of charged hadrons for krypton
($Kr$) and deuteron ($D$) targets as a function of \(p_{\perp}^{2}\) and as a function of \(z_{h}\)
at positron beam energy 27.6 GeV. The solid points correspond to HERMES
data \cite{Airapetian:2003mi} and the open points are obtained by HARDPING 2.0}
\label{fig:Kr}
\end{figure*}
\begin{figure*}[h!]
\centering
\vspace*{0.0cm}
\includegraphics[width=0.9\textwidth, height=0.3\textheight]{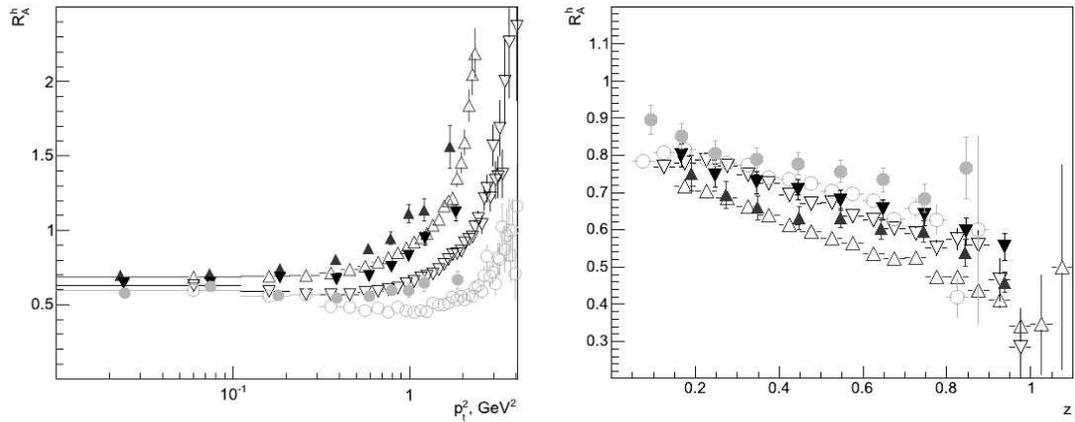}
\caption{Multiplicity ratio (\(R_{M}^{h}\)) of \(\pi^{+}\) -mesons for xenon
 ($Xe$) and deuteron ($D$) targets as a function of
\(p_{\perp}^{2}\) for different $z_{h}$ values and as a function of
\(z_{h}\) for different \(\nu\) values at
positron beam energy 27.6 GeV. The solid points correspond to HERMES data
\cite{Airapetian:2011jp} and the open points are obtained by HARDPING 2.0}
\label{fig:Xe}
\end{figure*}

Simulations of lepton-nuclei collisions
obtained by HARDPING 2.0 were compared with HERMES data
\cite{Airapetian:2003mi,Airapetian:2011jp}. The results are shown on Figs. \ref{fig:Kr},\ref{fig:Xe}.

The performed simulations shown a reasonable agreement of MC model HARDPING 2.0 with
the experimental HERMES data \cite{Airapetian:2003mi,Airapetian:2011jp}. This allowed to fix model
parameters such as inelastic  quark-nucleon (pre-hadron-nucleon) 
cross-section \(\sigma=10\) mb and
string tension \(k = 1.7\) GeV/Fm. Comparision with EMC \cite{Ashman:1991cx} and
 SLAC \cite{Benhar:1997zz,Degtyarenko:1994tt} data shown also a good agreement and it will be presented elsewhere.

To summarise, the effects of the two-stage hadronisation and multiple soft 
interactions inside of nucleus for
produced hadrons and their  constituents  were implemented into MC generator HARDPING 2.0. 
 The developed MC generator HARDPING 2.0 is allowed to describe reasonably well
the HERMES data \cite{Airapetian:2003mi,Airapetian:2011jp}
on hadron production in positron-nucleus scattering at 27.6 GeV.

The authors thank S.L. Belostotsky for useful discussions on the HERMES data.
This work was supported in parts by the Ministry of Education of the Russian
Federation, under the contract No. 02.740.11.0572 of the Federal task program
``Research and educational community of innovative Russia'' for 2009-2013 and by
the RF President grant NS-3383.2010.2.

\end{document}